\DeclareRobustCommand{\baselinestretch{2}}

\documentclass[onecolumn,showpacs,preprintnumbers,amsmath,amssymb]{revtex4}

\usepackage{graphicx}
\usepackage{dcolumn}

\def\p1{\partial_1}
\def\p2{\partial_2}
\def\p3{\partial_3}

\begin{document}
\title{Traversal of pulses through negative ($\varepsilon$, $\mu$) materials}
\author{ Lipsa Nanda and S. Anantha Ramakrishna}
\affiliation{Department of Physics, Indian Institute of Technology,Kanpur 208016, India}



\begin{abstract}
 We study the traversal times of electromagnetic pulses across dispersive media with 
negative dielectric permittivity ($\varepsilon$) and magnetic permeability ($\mu$) parameters. First we investigate the transport of optical pulses through an electrical plasma and a negative refractive index medium (NRM) of infinite and semi-infinite extents where no resonant effects come into play. The total delay time of the pulse constitutes of the group delay time and the reshaping delay time as analyzed by Peatross et al \cite{peatross}. For evanescent waves, even with broadband width, the total delay time is negative for an infinite medium whereas it is positive for the semi-infinite case. Evidence of the Hartman effect is seen for small propagation distance compared to the free space pulse length. The reshaping delay mostly dominates the total delay time in NRM whereas it vanishes when $\varepsilon(\omega)=\mu(\omega)$.

           Next we present results on the propagation times through a dispersive slab.
While both large bandwidth and large dissipation have similar effects in smoothening out the resonant features that appear due to Fabry-P\'{e}rot resonances, large dissipation can result in very small or even negative traversal times near the resonant frequencies. We investigate the traversal and the Wigner delay times for obliquely incident pulses. The coupling of evanescent waves to slab plasmon polariton modes results in large traversal times at the resonant conditions. We also find that the group velocity mainly contributes to the delay time for pulse propagating across a slab with refractive index (n) = -1. The traversal times are positive and subluminal for pulses with sufficiently large bandwidths. 
\end{abstract}

%
\maketitle 
\section{Introduction}
The time for light to traverse through a dispersive medium is interesting and important.A popular measure for the delay time of pulses is the Wigner delay time\cite{wigner}
$\tau_w = \frac {\partial \phi}{\partial \omega}|_{\omega=\bar {\omega}}$,
i.e, the frequency derivative of the phase of the output wave evaluated at the carrier frequency $\bar {\omega}$. The Wigner delay time
becomes inaccurate for large pulse bandwidths or when there is a large deformation in the pulses.

For pulses and particularly broadband pulses, Peatross {\it et al.}\cite{peatross} showed that the 
arrival time of a pulse at a point $\mathbf{r}$ can be well described by a 
time average over the component of the Poynting vector $\mathbf{S}$ normal to a (detector) surface at $\mathbf{r}$ as

\begin{equation} 
\langle t\rangle_\mathbf{r} = \frac{\mathbf{u}\cdot\int_{-\infty}^\infty t 
\mathbf{S}(\mathbf{r},t) dt}{\mathbf{u}\cdot\int_{-\infty}^\infty 
\mathbf{S}(\mathbf{r},t) dt}.
\end{equation}
Here $\mathbf{u}$ is taken to be the unit vector along the normal to the given surface.
The time of traverse between two points ($\mathbf{r}_i, \mathbf{r}_f $) is 
equal to the difference of the arrival times at the two points, and 
was shown analytically to consist of two parts: a contribution by the 
spectrally weighted average group delay at the final point $\mathbf{r} _f$
\begin{equation}
\Delta t_G = \frac{\mathbf{u}\cdot\int_{-\infty}^\infty 
\mathbf{S}(\mathbf{r}_f,\omega) \left[ (\partial~\mathrm{Re}\mathbf{k}/
\partial \omega) \cdot \Delta \mathbf{r}\right] d\omega}{\mathbf{u}\cdot
\int_{-\infty}^\infty\mathbf{S}(\mathbf{r}_f,\omega) d\omega},
\end{equation}
and a contribution that could be ascribed to the reshaping of the pulse
\begin{equation}
\Delta t_R = {\cal T} \left[ \exp(-\mathrm{Im}\mathbf{k}\cdot\Delta \mathbf{r})
\mathbf{E}(\mathbf{r}_i,\omega)\right] -{\cal T} \left[\mathbf{E}(\mathbf{r}_i,
\omega)\right], 
\end{equation}
which is calculated with the spectrum at the initial point $\mathbf{r} _i$. Here the operator ${\cal T}$ is 
\begin{equation} 
 {\cal T}  \left[\mathbf{E}(\mathbf{r},\omega)\right] =  \frac{\mathbf{u}
\cdot\int_{-\infty}^\infty Re\left[-i\frac{\partial \mathbf{E}(\mathbf{r},\omega)}
{\partial \omega} \times \mathbf{H}^{\ast}(\mathbf{r},\omega)\right] d \omega}
{\mathbf{u}\cdot\int_{-\infty}^\infty\mathbf{S}(\mathbf{r},\omega) d\omega},
\end{equation}
which represents the arrival time of a pulse at a point $\mathbf{r}$ in terms of the spectral fields.
The Poynting vector is represented by $\mathbf{S} (\mathbf{r},\omega) \equiv Re \left[\mathbf{E} (\mathbf{r},\omega)\times \mathbf{H} ^{\ast} (\mathbf{r}, \omega)\right]$. Here we take the real parts of the quadratic terms since we use complex representation for the fields, i.e., $e^{i(\mathbf{k} \cdot \mathbf{r}-\omega t)}$ for a plane wave.

The paper is organised into the following sections: The delay times for 
evanescent pulses and pulse transport through a plasma (infinite and semi-infinite with 
a boundary) are presented in Section-2. The delay times for pulse propagation 
in NRM (infinite and semi-infinite with a boundary) are discussed in section-3. The traversal times across a dispersive slab are discussed in section-4 and we conclude in Section-5 with a discussion of our results and their implications.

 \section{Arrival times for evanescent waves}
We will consider the arrival times for pulses composed entirely of 
evanescent waves. This is analogous to quantum mechanical tunneling of a 
particle under a barrier. Such situations arise directly in the transport
of radiation across a metal slab or under conditions of total internal 
reflection. 
Now consider the complex wave-vector in a medium, 
\begin{equation}
k^2 = \varepsilon \mu \frac{\omega^2}{c^2}.
\end{equation} 
In the limit of small imaginary parts of $\varepsilon$ and $\mu$, one can write
\begin{eqnarray}
k_r = \mathbf{Re}(k) \simeq \sqrt{\varepsilon_r\mu_r - 
\varepsilon_i\mu_i} \frac{\omega}{c}  \\
k_i = \mathbf{Im}(k) \simeq \frac{\varepsilon_r\mu_i + \varepsilon_i\mu_r}{
2 \sqrt{\varepsilon_r\mu_r -\varepsilon_i\mu_i}} \frac{\omega}{c},
\end{eqnarray}
where the subscripts $r$ and $i$ indicate the real and imaginary parts of the 
quantities. Thus for propagating waves, the real part of the wave vector depends primarily on
$\varepsilon_r$ \& $\mu_r$ while the imaginary part is directly proportional to
$\varepsilon_i$ \& $\mu_i$ or the dissipation. This however becomes
different for evanescent waves. To make clear the discussion for evanescent waves, we will 
consider an absorbing electric plasma with $\varepsilon_r < 0$, 
$\varepsilon_i >0$ and $\mu = \mu_r$. Now, 
\begin{eqnarray}
k_r \simeq \frac{1}{2} \sqrt{\frac{\mu_r}{\vert \varepsilon_r \vert}}
 \varepsilon_i  \frac{\omega}{c}, \\
k_i \simeq \sqrt{\vert\varepsilon_r\vert \mu_r} \frac{\omega}{c}.
\end{eqnarray}
Thus the real part of the wave-vector depends on the levels of dissipation in 
the medium ($\varepsilon_i$) and the imaginary part of the wave-vector which 
determines the decay of the wave depends on $\vert \varepsilon_r\vert$. This
implies, in turn, that the definitions of the group delay time and the 
deformation delay time given by Eqns. (2) and 
(3) respectively get interchanged for evanescent waves. This is an 
important difference for the arrival times of evanescent waves from 
that of propagating waves.

\subsection{Pulse traversal in an unbounded plasma}
Now we will investigate the arrival times for evanescent pulses inside a plasma.
For our calculations we will use a dielectric medium with the relative dielectric permittivity of the plasma to be of the causal form
\begin{equation}
\varepsilon(\omega)=1-
\frac{\omega_p^2}{\omega(\omega +i \gamma)},
\end{equation}
 and the relative magnetic permeability $\mu =1$.

 Now consider a pulse of light initially at ${r} _i$
whose electric field in time is given by 
\begin{equation}
\mathbf{E}({r} _i,t) = \hat{x} \mathbf{E} _0\exp[-\frac{t^2}{\tau^2}] \exp(-i\bar{\omega}t).
\end{equation}
We note that the product $\bar{\omega} \tau$, where $\bar{\omega}$ is the carrier frequency and $\tau$ is the pulse duration in Eq. (11) gives a criterion for broadband or narrowband pulses. Typically we take $\bar{\omega} \tau=$ 1000 or 100 for narrowband pulses and $\bar{\omega} \tau=$ 10 for broadband pulses.

 Here we consider that both the source of the radiation and the detector are embedded
inside the plasma and that the plasma is
unbounded. This is to avoid any effects of scattering from the boundaries and study the inherent effects of the 
plasma on the traversal times. The distance between the source and the detector
is taken to be $\Delta \mathbf{r}$.

 In Figs. 1(a) and 1(b), we plot the delay times obtained for pulse traversal inside the
unbounded plasma with a plasma frequency
$\omega_p= 10\gamma$. First of all, we note that the total delay time is negative for a large range of frequencies and mostly
superluminal below the plasma frequency. The total delay time is also dominated by the reshaping delay time at frequency
$\bar{\omega}$ less than $\omega_p$. Inside an infinite plasma, the absorption determines the energy transport, and hence the delay time as well. 
For larger bandwidths, the contribution of the group delay becomes appreciable at higher frequencies ($\bar{\omega} \sim 7\gamma$ to $10\gamma$ in Fig. 1(b)). 
We also note that the qualitative behaviour of the delay time
does not change appreciably with increase in distance between the source and the 
detector except for the difference in scales \cite{lipsa}.
\begin{figure}[tb]
\begin{center}
\includegraphics[angle=-0,width=0.5\columnwidth]{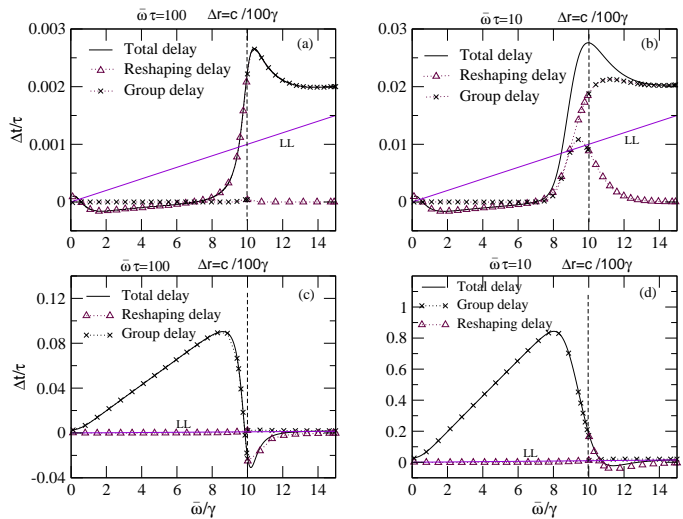}
\end{center}
\caption {The total delay time, the reshaping delay time and the group delay time represented
by  --, $\Delta$ and $\times$ as a function of the carrier frequency $\bar \omega$ 
(a)Delays for narrowband pulses ($\bar \omega \tau = 100$) and  $\Delta \mathbf{r} = $\^{z}$c/100\gamma$ in an unbounded plasma.
(b) Same as (a), but for a broadband pulse ($\bar \omega \tau = 10$). 
Frames (c) and (d) correspond to a semi-infinite plasma with the respective parameters similar to (a) and (b).
The vertical line drawn at  $\bar \omega/\gamma=10$ separates the propagating waves from the evanescent waves.
Note that the group delay and the reshaping delay times interchange their roles for the evanescent waves in comparison
to the propagating waves. The straight line going across the graphs denoted as LL, is the light line 
for free space propagation ($\Delta t = \Delta r / c$).}
\label{g1_spie}
\end{figure}

\subsection{Pulse traversal through a bounded plasma}
 Here we consider the situation of a plasma with a semi-infinite extent. We consider the
source of the radiation is inside the plasma at a distance $\Delta \mathbf{r}$ from the planar interface with vacuum. The detector is taken to be
in vacuum just outside the interface. This is more physical because there would be an interface (\textrm{impedance mismatched})
involved with the detector anyway. There is not coupling to any surface plasmon modes of the plasma-vacuum interface. The interface has an important effect
of changing the amount of energy that reaches the detector via the transmittance of the interface given by
\begin{equation}
T_p(\omega) = \frac{2\sqrt{\varepsilon_f}}{\sqrt{\varepsilon_p} + \sqrt{ \varepsilon_f}},
\end{equation}
where $T_p(\omega)$ is the Fresnel transmission coefficient
for the P- polarized light. $\varepsilon_f = 1$ is the relative dielectric permittivity in free space.
Thus the field at the detector is given by
\begin{equation}
\mathbf{H} (\mathbf{r}_f, \omega) = T_p(\omega)e^{i\mathbf{k} \cdot \Delta \mathbf{r}}  \mathbf{H} (\mathbf{r}_i , \omega),
\end{equation}
where $\mathbf{k}$ is the wave vector in the plasma.
 
 In Figs. 1(c) and 1(d), we plot the delay times for pulses from a source within a plasma
 for different bandwidths and with a source to interface distance $\Delta \mathbf{r} =$
\^{z}$c/100 \gamma $. First of all, we note that the reshaping delay time is negligible
compared to the group delay
time even for $\omega < \omega_p$. \textrm{The group delay} dominates the total delay time for small  $\Delta \mathbf{r} ($ \^{z}$c/10 \gamma $ and \^{z}$c/100 \gamma $).
This important difference from the case of an unbounded plasma results because the presence of the boundary causes
a reflected evanescent wave. Now energy transport is primarily determined by the phase difference of the incident evanescent wave and the reflected wave, and not only by the dissipation in the medium. Thus the group delay time
plays the determining role. Secondly it should
be noted that the total delay time is almost always positive except near the plasma frequency for narrowband pulses
and small $\Delta \mathbf{r}$ (\^{z}$c/10 \gamma$ and \^{z}$c/100 \gamma$).
One notes that negative delay times result for narrowband pulses ($\bar {\omega} \tau = 100$)
and small $\Delta \mathbf{r}$ even for propagating waves ($\varepsilon > 0$)
near $\bar {\omega} = \omega_p$. This is a consequence of the $\mathbf{k}=0$ mode at $\varepsilon = 0$. This 
negativity goes away for larger bandwidths.

 At much larger source to interface distances ($\Delta \mathbf{r} = $\^{z}$c/\gamma$),
the deformation of the pulse 
contributes appreciably to the total delay time. We show the delay times in Figs. 2(a) and 2(b) for
$\Delta \mathbf{r} = $\^{z}$ c/\gamma$ at different bandwidths. The behavior of the reshaping delay time
($\bar \omega < \omega_p$) tends to that of the behavior in an infinite plasma while the group delay time strongly 
moderates this contribution to the total delay, and the total delay time is positive almost everywhere. 
Surprisingly we note that there is a small region of 
frequencies where the total delay time goes negative even for broadband pulses, and at this large distance 
involved. However, we note that the spectral width of the region where the total time 
becomes negative, reduces with increasing pulse band width. Hence in the limit 
of very large bandwidths, we expect this spectral width to go to zero asymptotically.

\begin{figure}[tb]
\begin{center}
\includegraphics[angle=-0,width=0.5\columnwidth]{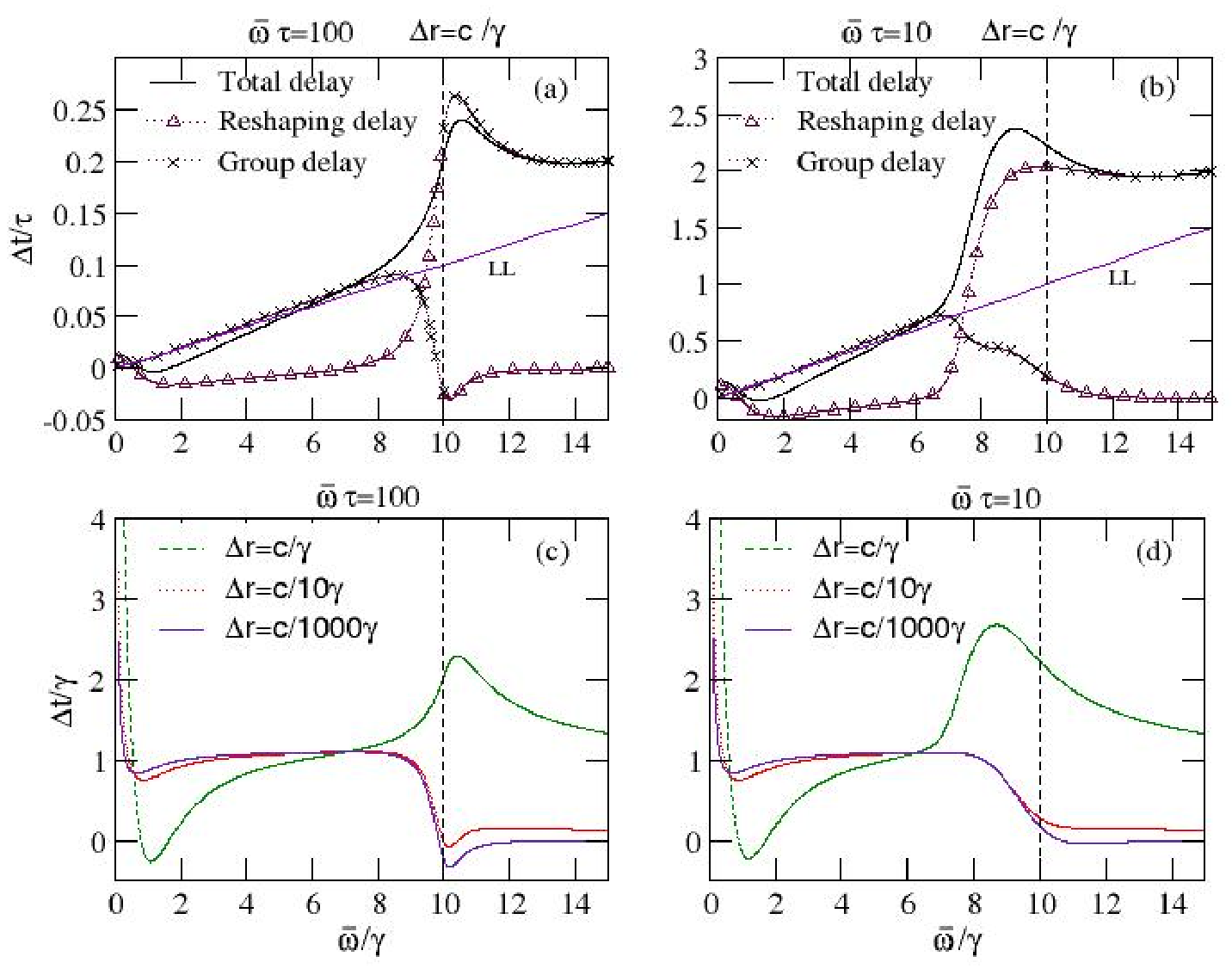}
\end{center}
\caption {(a) The various delay times plotted for large source-boundary distance ($\Delta \mathbf{r} = $\^{z}$c/\gamma$) and
narrowband pulses ($\bar \omega \tau = 100$) for a bounded semi-infinite plasma. The symbols are similar to Fig. 1.
(b) Same as (a), but for broadband pulses ($\bar \omega \tau = 10$). The Hartman effect can be observed in graphs
(c) and (d) which correspond to the total delay times for various parameters shown.}
\label{gset4}
\end{figure}

 We note the presence of a Hartman effect in our calculations as well. In Figs. 2(c) and
2(d), we plot the delay times
with respect to the carrier frequency without scaling with respect to the temporal pulse width. We find that over a large 
range of carrier frequencies below the plasma frequency, the delay time is almost the same for various distances involved
($\Delta \mathbf{r} = $\^{z}$c/10 \gamma$ and \^{z}$c/1000 \gamma$).
This is seen for both broadband pulses as well as narrowband pulses for these $\Delta \mathbf{r}$.
This saturation of the delay time with distance is a generalization of the Hartman effect for 
broadband pulses which is usually noted for
monochromatic evanescent waves. However for much larger distances ($\Delta \mathbf{r} = $\^{z}$c/\gamma$) which are
comparable to the spatial pulse width in free space ($l = c \tau$), the deformation takes over and the Hartman effect
is lost.

\section{Arrival times in negative refractive index medium}
In this section, we will discuss the arrival times for pulses propagating through negative refractive index medium with infinite and semi-infinite extent.
A sufficient condition for negative refractive index is $\varepsilon < 0$ and $\mu < 0$ at any frequency.

An unusual situation arises when the value of $\varepsilon$ becomes equal to that of $\mu$ (a case of propagating waves). For this case, the reshaping delay turns out to be zero. Note that
the value of the refractive index becomes equal to $\varepsilon$ or $\mu$ when $\varepsilon = \mu$.
The electric field at the initial point(in the frequency domain) is given by
\begin{equation}
\mathbf{E} (\mathbf{r} _i,\omega) = \hat{x} \frac{\mathbf{E} _0}{2\sqrt{2}}\tau 
e^{-\frac{(\omega - \bar {\omega})^2}{4}\tau^2}
\end{equation} 
which is the same field as given by Eq. (11), but calculated in the frequency domain.
In an unbounded medium, the electric field at the final point is related to that at the initial point as
\begin{equation}
\mathbf{E} (\mathbf{r} _f,\omega) = \mathbf{E} (\mathbf{r} _i,\omega)e^{i\mathbf{k} \cdot \Delta \mathbf{r}},
\end{equation}
where $\mathbf{r} _f = \mathbf{r} _i + \Delta \mathbf{r}$ and $\mathbf{k}$ is the wave vector in the medium.
The magnetic field is related to the electric field through the Maxwell's equation.
\begin{equation}
\mathbf{H} (\mathbf{r} _i,\omega) = \hat{y} \frac{\mathbf{E} _0}{2\sqrt{2}}\frac{1}{c\mu _0}\tau 
e^{-\frac{(\omega - \bar {\omega})^2}{4}\tau^2},
\end{equation}
and
\begin{equation}
\mathbf{H} (\mathbf{r} _f,\omega) = \hat{y} \frac{\mathbf{E} _0}{2\sqrt{2}}\frac{1}{c\mu _0}\tau
e^{-\frac{(\omega - \bar {\omega})^2}{4}\tau^2} e^{i\mathbf{k} \cdot \Delta \mathbf{r}}.
\end{equation}
Using these, we calculate the delay time for pulse propagation between the initial and the final positions
and see that the total delay time consists of only one non-zero term which is the group delay time,
\begin{equation}
\Delta t = \frac{\int_{-\infty}^\infty
\tau ^2e^{-\frac{(\omega - \bar {\omega})^2}{2}\tau^2}e^{-2\mathrm{Im} \mathbf{k} \cdot \Delta \mathbf{r}}
\left[ (\partial~\mathrm{Re}\mathbf{k}/
\partial \omega) \cdot \Delta \mathbf{r}\right] d\omega}{\int_{-\infty}^\infty\tau ^2e^{-\frac{(\omega - \bar {\omega})^2}{2}\tau^2}e^{-2\mathrm{Im} \mathbf{k} \cdot \Delta \mathbf{r}}
 d\omega}.
\end{equation}
This means that the reshaping delay time identically vanishes and
no reshaping of the pulse takes place in a medium with $\varepsilon = \mu$. Even for a bounded medium with an interface separating the given medium from vacuum, due to perfect impedance matching, the transmission coefficient through the interface is unity. This means that the dispersion in the transmission of the pulse through the interface plays no role and the reshaping delay disappears here too.

We take the Drude Lorentz form for
$\varepsilon$ given by Eq. (10) and a Lorentz dispersion for $\mu$ given by
\begin{equation}
\mu = 1 + \frac{\omega_m^2}{\omega_{0m}^2 - \omega^2 - i\omega \gamma}.
\end{equation}

For convenience, we take $\omega_{0m} = 5\gamma$, $\omega_m^2 = 64\gamma ^2$
and the rate of dissipation $\gamma$ in $\mu$ to be the same as that in $\varepsilon$.
This results in an electric plasma ($\varepsilon < 0$, $\mu > 0$) for $0 < \omega < 5\gamma$,
a negative refractive index medium ($\varepsilon < 0$, $\mu < 0$) for $5\gamma < \omega < 9.434\gamma$,
an electric plasma ($\varepsilon < 0$, $\mu > 0$) for $9.434\gamma < \omega < 10\gamma$ and a positive 
refractive index medium ($\varepsilon > 0$, $\mu > 0$) for $\omega > 10\gamma$.

 We study the behaviour in an infinite medium.
We plot the delay times(total, group and reshaping delays) for a pulse in Fig. 3. We note that at low frequencies,
the delay time is negative as in an infinite plasma. However, there is a large peak in the total delay time at
$\bar {\omega}=\thickapprox{3\gamma}$. The group delay almost exclusively contributes to this.
This is presumably due to the rapid increase in Re($\mu$) (as seen from the dispersion curve). In the negative refractive region, when there are propagating
waves in the medium, there is appreciable contribution from both group and reshaping delays.
The reshaping delay time is mostly negative. We note that the total delay time is always positive as well as subluminal in the negative index frequency region.
\begin{figure}[tbp]
\begin{center}
\includegraphics[angle=-0,width=0.5\columnwidth]{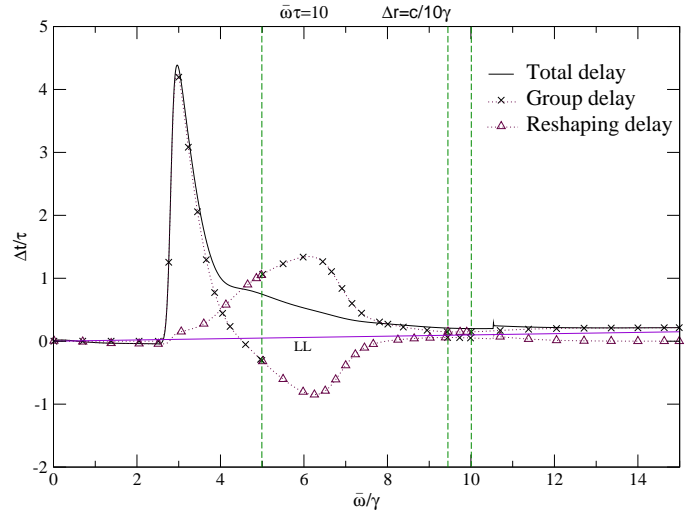}
\end{center}
\caption {The various delay times plotted as a function of $\bar \omega$ in an unbounded medium 
which behaves like a plasma and with negative or positive refractive indices for 
certain frequency ranges for broadband pulses ($\bar \omega \tau = 10$) and $\Delta \mathbf{r} = $\^{z}$c/10 \gamma$.
The vertical lines at frequencies $5\gamma$, $9.434\gamma$ and $10\gamma$ separate the 
frequency regions where the field modes are either evanescent or propagating. The symbols are similar to Fig. 2.}
\label{gset6}
\end{figure}

\begin{figure}[tb]
\begin{center}
\includegraphics[angle=-0,width=0.5\columnwidth]{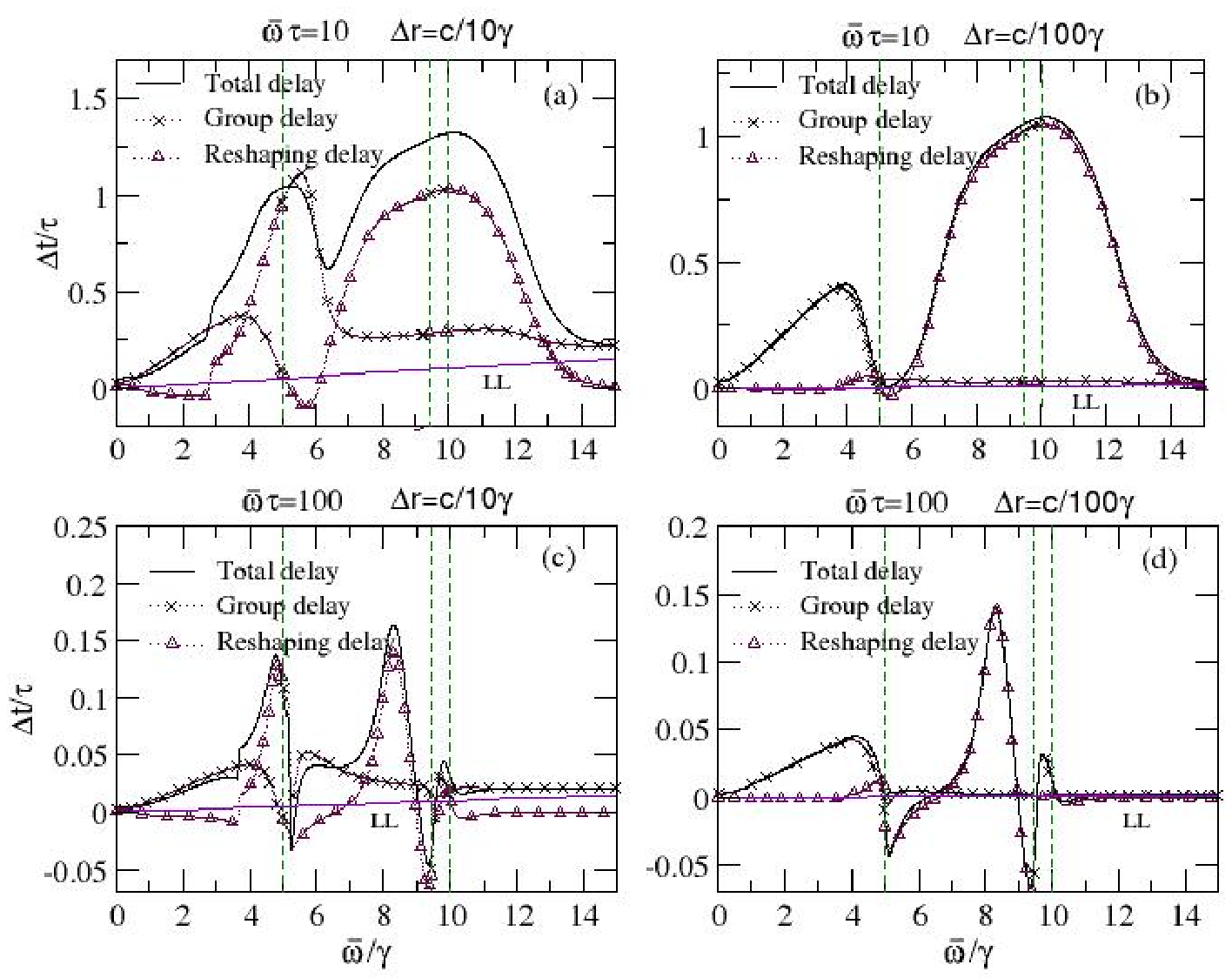}
\end{center}
\caption {(a) The various delay times plotted as a function of $\bar \omega$ in a bounded medium. 
The parameters and symbols are similar to Fig. 3. (b) Same as (a), with $\Delta \mathbf{r} = $\^{z}$c/100 \gamma$.
Frames (c) and (d) correspond to narrowband pulses ($\bar \omega \tau = 100$) with source-boundary distances
$\Delta \mathbf{r} = $\^{z}$c/10 \gamma$ and $\Delta \mathbf{r} = $\^{z}$c/100 \gamma$.}  \label{gset7}
\end{figure}

 Now, we take a semi-infinite medium with the above material dispersion.
As in section 2.2, we take the source to be inside the semi-infinite medium at a distance $\Delta \mathbf{r}$
from the interface with vacuum and detect the radiation outside.
As before, we consider only waves with zero parallel wave vector($\mathbf{k}_\parallel = 0$). We note that the reshaping delay 
plays a major role in determining the times in the negative refractive
index region. In fact for broadband radiation, bulk of the total delay time comes from the 
reshaping delay (Figs. 4(a) and 4(b)). For narrowband pulses ($\bar \omega \tau = 100$), the total delay time
can become negative near regions where a change of sign occurs in $\mu$ or $\varepsilon$ (Figs. 4(c) and 4(d)) .
But for broadband pulses ($\bar \omega \tau = 10$), this negativity disappears even for short distances of propagation.

\section{Traversal time across a dispersive slab}

Next we calculate the propagation times for electromagnetic pulses across a dispersive slab, the dielectric permittivity ($\varepsilon$) and the magnetic permeability ($\mu$) of which are given by Eqs. (10) and (19) with respective damping constants equal to $\gamma_p$ and $\gamma_m$.
\begin{equation}
\varepsilon(\omega)=1-
\frac{\omega_p^2}{\omega(\omega +i \gamma_p)},
\end{equation}

\begin{equation}
\mu(\omega) = \frac{\omega^2-\omega_b^2+ i\omega \gamma_m}{\omega^2- \omega_{0}^2 + i\omega \gamma_m}.
\end{equation}
Here $\omega_b^2=\omega_0^2+\omega_m^2$ and
$\omega_{p,b,o}$=2$\pi f_{p,b,o}$ with $f_p$, $f_b$, and $f_0$ taken to be 12 GHz, 6 GHz and 4GHz respectively.
The slab behaves as a positive refractive index medium (PRM) or right handed medium (RHM) ($\varepsilon>0$, $\mu>0$)
when $\omega > \omega_p$, a negative refractive index
medium (NRM or LHM) ($\varepsilon<0$, $\mu<0$) within $\omega_0 < \omega < \omega_b$, and as a barrier ($\varepsilon<0$, $\mu>0$) elsewhere.
 We take the source of radiation to be placed in vacuum just outside one boundary of the
slab and the detector just outside the other one. We have taken same medium (vacuum) on either sides of the slab. We denote the sides of the source and the detector respectively by regions 1 and 3 and the slab by region 2.
So here $\varepsilon_{1}=\mu_{1}=\varepsilon_{3}=\mu_{3}=1$. $\varepsilon_{2}$, and $\mu_{2}$ are
respectively given by Eqs. (20), and (21). We take our initial pulse of the form given by Eq. (14).
The magnetic field is simply obtained using the Maxwell's equations,
\begin{equation}
\mathbf{H} (\mathbf{r} _i,\omega) = \hat{y} \frac{\mathbf{E} _0}{2\sqrt{2}}\frac{k_{z1}}{\omega\mu_1\mu _0}\tau
e^{-\frac{(\omega - \bar {\omega})^2}{4}\tau^2},
\end{equation}
Here $k_{z1}$ represents the wave vector in the first medium where the source is present.

For the P-polarization, the magnetic field at the detector is related to that at the source
via the transmission coefficient across the slab. The final magnetic and the electric fields at the detector are given by
\begin{equation}
\mathbf{H} (\mathbf{r} _f,\omega) = \hat{y} \frac{\mathbf{E} _0}{2\sqrt{2}}\frac{k_{z1}}{\omega\mu_1\mu _0}\tau
e^{-\frac{(\omega - \bar {\omega})^2}{4}\tau^2}\mathbf{T(\omega)},
\end{equation}
and
\begin{equation}
\mathbf{E} (\mathbf{r} _f,\omega) = \hat{x} \frac{\mathbf{E} _0}{2\sqrt{2}}\frac{k_{z1}k_{z3}}{\omega^2\mu_1\varepsilon_3}
c^2\tau e^{-\frac{(\omega - \bar {\omega})^2}{4}\tau^2}\mathbf{T(\omega)},
\end{equation}
where $k_{z3}$ represents the wave vector in the third medium.

Here $\mathbf{T(\omega)}$ represents the transmission coefficient across the slab which is given by,
\begin{equation}
\mathbf{T(\omega)} =\frac{tt' e^{ik_{z2}\Delta r}}{1-r'^2 e^{2ik_{z2}\Delta r}},
\end{equation}
where $\Delta r$ represents the slab thickness and $k_{z2}$ represents the wave vector inside the dispersive slab.
$t$, $t'$, and $r'$ respectively represent the Fresnel coefficients
of transmission, and reflection by the slab interfaces and are given by,
\begin{eqnarray*}
t=\frac{2\frac{k_{z1}}{\varepsilon_1}}{\frac{k_{z1}}{\varepsilon_1}+\frac{k_{z2}}{\varepsilon_2}},
t'=\frac{2\frac{k_{z2}}{\varepsilon_2}}{\frac{k_{z2}}{\varepsilon_2}+\frac{k_{z3}}{\varepsilon_3}},
r'=\frac{\frac{k_{z2}}{\varepsilon_2}-\frac{k_{z3}}{\varepsilon_3}}
{\frac{k_{z2}}{\varepsilon_2}+\frac{k_{z3}}{\varepsilon_3}}.
\end{eqnarray*}
Here the unprimed, and primed coefficients stand respectively for the coefficients across the first and the second boundaries of the slab.
For S-polarization, in the expressions of the Fresnel coefficients, the $\varepsilon$'s are simply replaced by $\mu$'s.

 Also the Fresnel coefficients relate the electric fields across the interface rather
 than the magnetic fields.
Suffixes 1, 2, and 3 respectively represent the parameters at the source, slab, and the detector sides as described earlier.
For convenience we later substitute equal material parameters on both (source and detector) sides of the slab.
We calculate the delay times for different bandwidths (Narrow and broadbands).

The Wigner delay time was calculated using Eq. (25), and is given by,
\begin{equation}
\tau_{\omega}=\frac{\partial \phi}{\partial \omega}=\frac{\frac{\partial p}{\partial \omega} \tan(k_{z2}\Delta r)+ p \sec^2(k_{z2}\Delta r)\frac{\partial k_{z2}}{\partial
\omega} \Delta r}{1 + p^2 \tan^2(k_{z2} \Delta r)},
\end{equation}
where
\begin{equation}
p = \frac{\frac{k_{z1}\varepsilon_2}{k_{z2}\varepsilon_1}+\frac{k_{z2}\varepsilon_1}{k_{z1}\varepsilon_2}}{2}
\end{equation}
for P-polarization and
\begin{equation}
p = \frac{\frac{k_{z1}\mu_2}{k_{z2}\mu_1}+\frac{k_{z2}\mu_1}{k_{z1}\mu_2}}{2}
\end{equation}
for S-polarization. Throughout our calculations (both for normal and oblique incidence), we have checked that the Wigner
delay time yields the same result as the traversal time for narrowband pulses with the average energy flow method.

\subsection{Traversal times for normal incidence}

\begin{figure}[tb]
\begin{center}
\includegraphics[angle=-0,width=0.50\columnwidth]{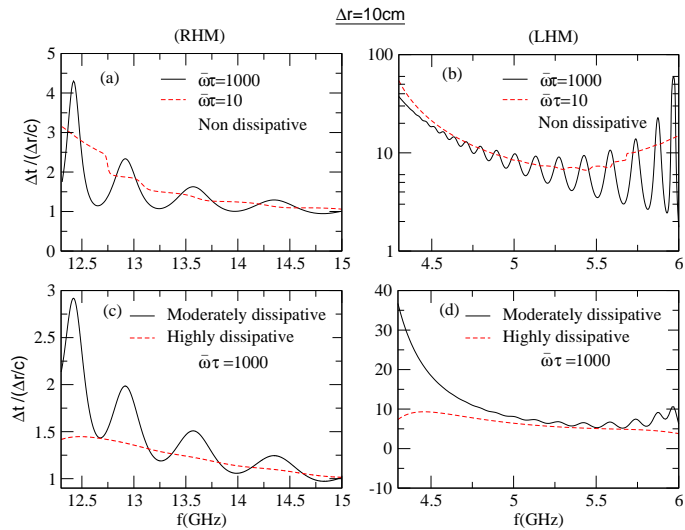}
\end{center}
\caption {(Color online) Scaled total delay time of a pulse with different bandwidths ($\bar{\omega} \tau$) plotted as a
function of the carrier frequency (f) across a dispersive slab with large thickness ($\Delta r =10cm$).
(a) Solid, and dashed lines show delay times respectively for a narrowband, and a broadband pulse across a
nondissipative slab of RHM.
(b) Same as (a), but across a slab of LHM.
(c) Delay time for a narrowband pulse across a moderately dissipative ($\gamma_p=0.01\omega_p$, $\gamma_m=0.01\omega_b$)
slab (solid line), and a highly dissipative ($\gamma_p=0.1\omega_p$, $\gamma_m=0.1\omega_b$)
slab (dashed line) of RHM. (d) Same as (c), but across a slab of LHM.}
\label{plot2}
\end{figure}

In this case, the parallel component of the wave-vector is zero and the pulse
is normally incident on the slab. So there is no coupling with the slab plasmon polaritons. Since $k_x=0$,
the Maxwell's equations can be combined to give,
\begin{equation}
k_z^2=\frac{\omega^2\varepsilon\mu}{c^2}.
\end{equation}
This is independent of whether the slab is of a RHM or a LHM.
In Fig. 5, we plot the delay times scaled with the free space propagation ($\frac{\Delta r}{c}$) versus the carrier frequency
for both broadband ($\bar{\omega} \tau$= 10) and narrowband ($\bar{\omega} \tau$= 1000) pulses.
 For narrowband pulses, we refer to \cite{duttagupta} where
the Wigner delay times were calculated for a nondissipative slab and it was shown that
resonant features appear in  the delay time behaviors due to presence of the poles of the transmission
coefficient (Fabry-P\'{e}rot resonances).
We have taken ($\Delta r =10$cm) as large thickness and ($\Delta r =1$cm) as small thickness
of the slab relative to the wavelength (2.5cm) of the pulse at the electrical plasma frequency ($f_p$) .
First we compute the results for the traversal times of broadband
and narrowband pulses through a nondissipative slab which is achieved by substituting
$\gamma_p = \gamma_m = 0$ in the expression of
$\varepsilon$ and $\mu$. Figs. 5(a) and 5(b) respectively show the traversal times for pulse propagation inside slabs with positive and negative refractive indices. In both the figures, it can be observed that,
the features due to the slab resonances get smoothened with an increase in the pulse bandwidth.
So it is expected that for extremely broadband light,
these features might completely disappear. Here we note that, the results for the narrowband pulses in Figs. 5(a), and 5(b), are exactly the same as those for the Wigner delay times given in \cite{duttagupta}.
Next we study the traversal times for narrowband pulses
propagating through dissipative slabs of both RHM and LHM (Figs. 5(c), and 5(d)).
To include moderate levels of dissipation in the medium,
we use $\gamma_p=0.01\omega_p$ and $\gamma_m=0.01\omega_b$ and for high levels of dissipation, we use
$\gamma_p=0.1\omega_p$ and $\gamma_m=0.1\omega_b$ respectively in Eqs. (20) and (21).
We see that when a small amount of dissipation is introduced in the medium, the time taken for transmission through the
slab is less than that taken for the nondissipative case. With increased dissipation in the
slab, one can also clearly observe that the slab resonant features disappear.

\begin{figure}[tb]
\begin{center}
\includegraphics[angle=-0,width=0.50\columnwidth]{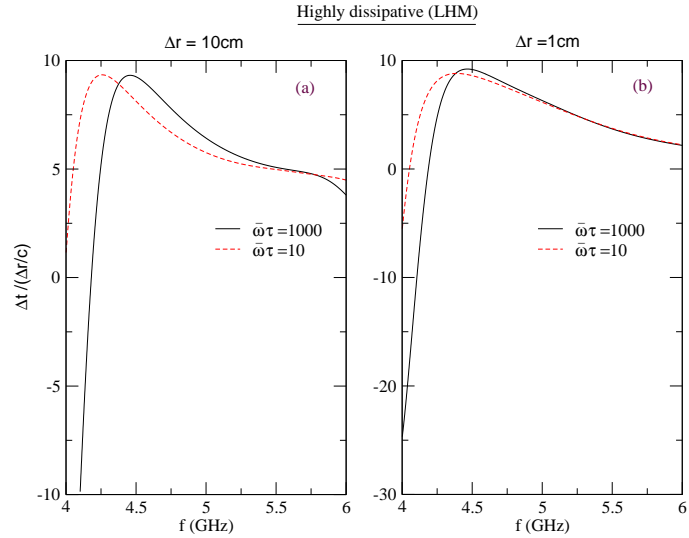}
\end{center}
\caption {(Color online) (a) Scaled total delay time as a function of the carrier frequency (f) for a
narrowband pulse (solid line), and a broadband pulse (dashed line) across a highly
dissipative ($\gamma_p=0.1\omega_p$, $\gamma_m=0.1\omega_b$) slab of LHM, and
large thickness ($\Delta r=10cm$). (b) same as (a), but with a small slab thickness ($\Delta r=1cm$).}
\label{plot3}
\end{figure}

We have also studied the delay times for a highly dissipative slab of
LHM for both narrow and broadband pulses (Figs. 6(a), and 6(b)).
Fig. 6(a) shows the results for a slab with large thickness (10cm) and Fig. 6(b) shows the corresponding results for a slab with a small
thickness (1cm). It can be clearly seen that the delay time is very small near the magnetic resonance frequency. Then it
rapidly increases for large frequencies, and after passing through a peak, it gradually decreases.
For narrowband pulses, the total delay time near the
resonance frequency ($\omega_{0}$) even becomes largely negative. Even for broadband pulses, with small thickness
of the slab, this negativity in the delay time appears near $\omega_{0}$ although to smaller and smaller
extent with increasing thickness of the slab.
The anomalous dispersion of the refractive index of a medium with high amount of dissipation leads to
small/negative delay times for broadband/narrowband pulses near the resonance frequency.

\subsection{Traversal times for oblique incidence}

\begin{figure}[tb]
\begin{center}
\includegraphics[angle=-0,width=0.50\columnwidth]{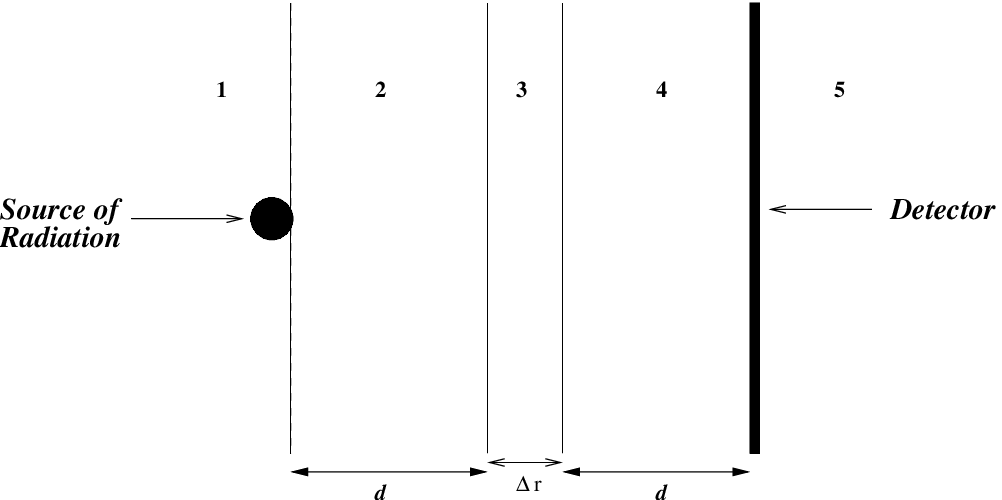}
\end{center}
\caption {A layered structure consisting of five slabs used to study the tunneling of pulses.
Region 3 is the dispersive slab with thickness $\Delta r$
with $\varepsilon_3$, and $\mu_3$ respectively given by Eqs. (20), and (21). Regions 2, and 4 are air slabs
($\varepsilon_2=\mu_2=1$), and with thickness d. Regions 1, and 5 are semi-infinite slabs with
$\varepsilon_1=\varepsilon_5=25$, and $\mu_1=\mu_5=1$. The black circle in Region 1 just outside the boundary of slab 2 represents the source, and the black screen in Region 5 outside slab 4 represents the detector.}
\label{plot7}
\end{figure}

Next, we study the traversal time of a pulse when it is obliquely incident on a slab, i.e., the
parallel wave vector ($k_x$) is nonzero. For this case, the Maxwell equations give,
\begin{equation}
k_z=\sqrt{\frac{\omega^2\varepsilon\mu}{c^2}-k_x^2}=\frac {\omega}{c}\sqrt{\varepsilon \mu -q_x^2},
\end{equation}
where $k_x=q_x\frac{\omega}{c}$.

We consider the incident pulses with either S or  P polarizations. For convenience, we have scaled all the
frequencies with respect to the plasma frequency. So here $f_p=1$, $f_0=0.33$, and $f_b=0.5$.
The thicknesses of the slab used are $\Delta r=\frac{\lambda_p}{5}$ (small thickness), for which
$k_p \Delta r = 0.4\pi$, and $\Delta r= 2{\lambda_p}$ (large thickness), for which $k_p \Delta r = 4\pi$.
The thicknesses are relative to the wavelength ($\lambda_p$) at the plasma frequency.
Here we consider pulses for which most of the wave vectors are imaginary making the incident wave evanescent.
This is achieved by making the second term in Eq. (30) under the square root larger than the first term.
We calculated both the Wigner delay time and the Energy delay time for such pulses most of whose components
are evanescent.
While the Wigner delay time can be easily calculated using the phase shifts, calculating the traversal time via the
energy flow for evanescent pulses is a non-trivial problem. This is because the energy flow associated with a single
evanescent wave in vacuum is zero. Thus the Poynting vector for pulse with all spectral components having evanescent
wave vectors is zero at the detection point in vacuum. Hence it is not possible to compute the energy traversal time for evanescent pulses if the slab is embedded in vacuum. One needs to couple the energy in these systems out to measure
the pulses. For this, we make an arrangement with
layered slabs where the evanescent waves are outcoupled to propagating modes in high-index media.

In this arrangement, we have taken two nondispersive slabs of different parameters kept
symmetrically on either sides of the dispersive slab making a layered slab structure (Fig. 7).
The first and fifth slabs have semi-infinite extent
with large relative dielectric permittivity ($\varepsilon = 25$) and relative magnetic permeability ($\mu=1$).
The second and the 4th slabs consist of vacuum with $\varepsilon=1$, and $\mu=1$ and large slab thickness (d)
with corresponding $k_{p}d$ equal to (4$\pi$).
The 3rd or the middle dispersive slab has $\varepsilon$, and $\mu$ respectively given by Eqs. (20), and (21) and small slab
thickness ($\Delta r$) with corresponding $k_{p}\Delta r$ equal to (0.4$\pi$).
The source is present in the first medium just outside the boundary of the second slab and the detector is placed
in the fifth medium just outside the boundary of the 4th slab.
The value of $q_x$ in Eq. (30) is chosen in such a manner that the wavevector is real, making the pulse
propagating in first and fifth slabs, and imaginary making the pulse evanescent in second and 4th slabs.

First, we plot the Wigner delay time versus both the frequency and the wave vector in a moderately dissipative slab for P-polarization.
In a rather uniform landscape of delay times, the resonant conditions for the slab surface plasmon polaritons (SPPs)
stand out in stark contrast where the magnitude of the delay times are comparatively very large. Thus the entire
dispersion of the SPPs can be traced out (Fig. 8). There are two distinct plasmon modes corresponding to the
symmetric and antisymmetric modes whose frequencies tend to
 $\frac{f_p}{\sqrt{2}}$ at large wave vectors. Similarly two modes also appear below the
 magnetic resonant
frequency. For highly dissipative slabs also, such plasmon modes are seen for
evanescent waves, but with large broadening of the dispersion curves (Fig. 8(b)).

The surface plasmon features are lost when the thickness of the slab is larger than $\lambda_p$.

Then we studied the traversal times for evanescent pulses having extremely narrow bandwidth ($\bar{\omega} \tau = 5000$)
using the energy transport method with our new arrangement of the layered slab structure (Fig. 7).
For this arrangement, we plotted the delay times for narrowband pulses versus both the frequency and the wavevector,
and analysed the results for moderate and large dissipative slabs with both P-polarization
(Figs. 8(c), and 8(d)) and S-polarization.
We see that Figs. 8(c), and 8(d) look almost same as Figs. 8(a), and 8(b).
Thus, the energy traversal times are also significantly affected at the Surface Plasmon Polariton frequencies.
It is worth noting that the traversal times are large at the resonant conditions.

\subsection{Propagation through a slab with $n=-1$}
Finally, we consider a slab having unit negative refractive index ($n=-1$) and surrounded by vacuum ($n=+1$).
Negative refractive index of unit magnitude can be achieved at a single frequency for a nondissipative slab.
The properties of such a slab with ($n=-1$) are very interesting due to the possibility of designing
a perfect lens \cite{pendry,sar05}
By choosing $f_{p}=1$, $f_{0}=0.33$, $f_{b}=0.5$, we get $n=-1$ at $f=\frac{f_p}{\sqrt{2}}$ ($\varepsilon$=-1, $\mu$=-1).
With propagation inside the medium, the propagation distance increases by a factor of $\frac{1}{\cos\theta}$.
Here due to perfect impedance matching, no multiple reflections take place.
Using the expression for group velocity ($v_g=\frac{\partial \omega}{\partial k}$),
the group delay along the direction of propagation is given by
$G_d=\frac{\Delta r}{v_g\cos\theta}=\frac{\Delta r}{c\cos\theta} (n+\omega\frac{\partial n}{\partial \omega})$.
For the particular frequency $f=\frac{f_p}{\sqrt{2}}$, the second term within the bracket in the above expression gives a value equal
to $\frac{32}{7}$. We plot both the Wigner delay time, and the group delay time versus $q_x$ for propagating pulses for
the particular frequency mode described above (Fig. 9). From the graph, it can be observed
that the delay time gradually increases with $q_x$ until  $q_x=1$  (where it becomes infinity).
It can be seen that the graph feature of the group delay time is very similar to the Wigner delay time.
Hence it is  inferred that the group delay mainly contributes to the total delay occuring
during the propagation of a pulse inside a slab with  $n=-1$.

\begin{figure*}[tb]
\begin{center}
\includegraphics[angle=-0,width=0.85\columnwidth]{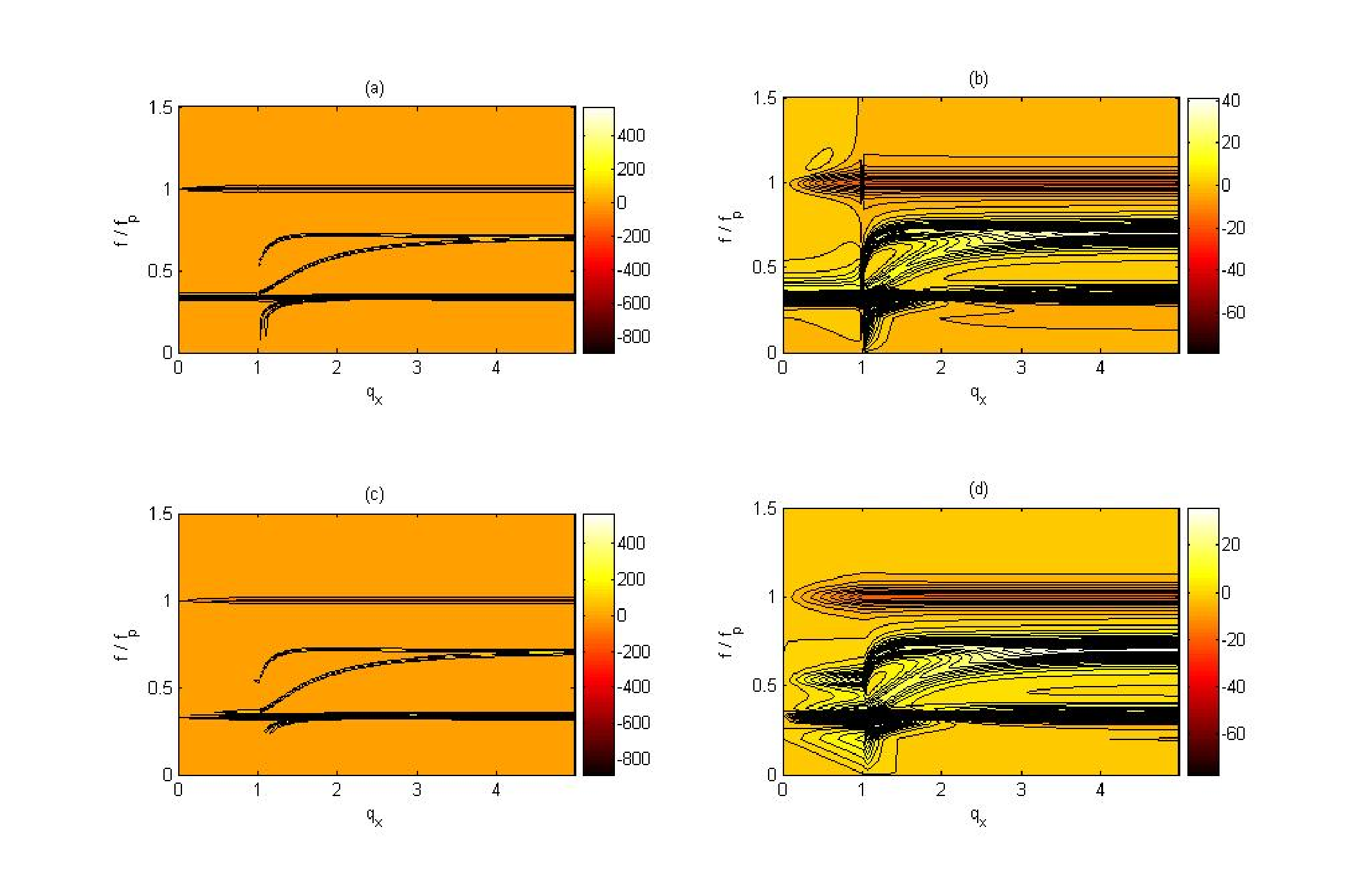}
\end{center}
\caption {(Color online) (a) The Wigner delay times for pulse traversal with P-polarization as a function of
scaled carrier frequency ($f/f_p$) and scaled parallel wave vector ($q_x = \frac{k_{x}c}{\omega}$)
across a moderately dissipative ($\gamma_p=0.01\omega_p$, $\gamma_m=0.01\omega_b$)
slab with small thickness ($\Delta r=\frac{\lambda_p}{5}$). (b) Same as (a), but for a highly
dissipative ($\gamma_p=0.1\omega_p$, $\gamma_m=0.1\omega_b$) slab. (c)The traversal times obtained by the energy transport method across the layered
slab structure where the middle dispersive slab is moderately dissipative ($\gamma_p=0.01\omega_p$, $\gamma_m=0.01\omega_b$)
in nature with small thickness ($\Delta r=\frac{\lambda_p}{5}$).
(d) same as (c), but for a highly dissipative ($\gamma_p=0.1\omega_p$, $\gamma_m=0.1\omega_b$) slab.
The dispersion of the slab plasmon polariton modes of the slab
stand out clearly and the resonant conditions for these modes are characterized by large Energy delay times.}
\label{plot8}
\end{figure*}

\section{Conclusions}
In summary, we have shown that the very definitions of the average group delay time and the reshaping delay time for the evanescent pulses get interchanged. We have also shown that in an infinitely
extended plasma, the delay time is primarily determined by the reshaping
delay time and is usually negative. In a bounded plasma when the
radiation is detected outside the boundary (in vacuum), the group delay time
dominates and the total delay times are usually positive and subluminal
for large enough frequency bandwidths associated with the pulses. We also note the Hartman effect in the context of energy transport for
evanescent pulses when the source to boundary distance is small compared to
the free space pulse length. In the case of negative refractive index materials, the total delay times are dominated by the reshaping delay times.

For a pulse traversing across a dispersive slab, we have shown that a high amount of dissipation in the slab material, along with large pulse bandwidth, smoothen out the resonant features. We have analysed the reason behind the occurrence of small/negative delay times near the
magnetic resonant frequency which is a consequence of
anomalous dispersion of the refractive index of the slab medium. We have also shown that the group delay mainly contributes to the
total delay across a slab with unit negative refractive index, and surrounded by vacuum.

\begin{figure}[tb]
\begin{center}
\includegraphics[angle=-0,width=0.5\columnwidth]{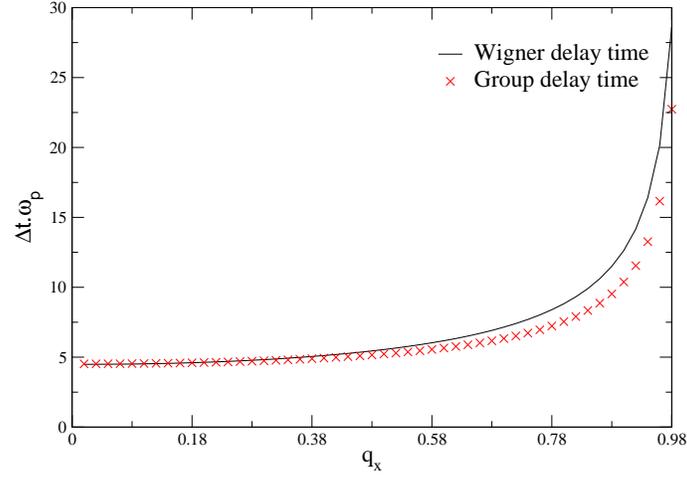}
\end{center}
\caption {(Color online)The Wigner delay time (solid line), and the Group delay time (cross symbol) plotted versus
scaled parallel wave vector ($q_x = \frac{k_{x}c}{\omega}$) across a slab with small thickness
($\Delta r=\frac{\lambda_p}{5}$), and unit negative refractive index ($n=-1$) surrounded by vacuum.}
\label{plot9}
\end{figure}

\section*{Acknowledgement}
SAR acknowledges support from the Department of Science and Technology, India under grant no.SR/S2/CMP-54/2003. LN acknowledges her fellowship from the University Grants Commission, India.

\end{document}